# Multi-Agent Model using Secure Multi-Party Computing in e-Governance

Dr. Durgesh Kumar Mishra, Samiksha Shukla

**Abstract** — Information management and retrieval of all the citizen occurs in almost all the public service functions. Electronic Government system is an emerging trend in India through which efforts are made to strive maximum safety and security. Various solutions for this have been proposed like Shibboleth, Public Key Infrastructure, Smart Cards and Light Weight Directory Access Protocols [1]. Still, none of these guarantee 100% security. Efforts are being made to provide common national identity solution to various diverse Government identity cards. In this paper, we discuss issues related to these solutions.

—————————— ◆ ——————————

## 1 INTRODUCTION

Digitalization of all the Government activities and offices has given rise to the concept of e-Governance which is vital to the IT revolution. It has to be done keeping in mind the large diverse cultural background. Fast progress in network technologies and development in data mining and distributed data applications have encouraged the computerization in even Government sector, which was till now untouched.

Privacy Preserving data mining, secure multiparty computation, cryptography, randomization and anonymization can be the suggestive security mechanisms [8] that needs to be enforced for all the Government information systems. Data of such Government offices needs to be kept extremely secure and confidential with proper accessing and authorization checks of individual employee.

With an e-Governance System, we try to provide a common information system that can maintain and serve the common information requirements of various diverse public and private sectors. This involves large data transfer among various organizations and hence, citizen privacy is at stake, if not properly checked and administered. Special attention needs to be provided for such web-based applications. E-governance, as a concept, can be a great technological step ahead. It can be implemented using privacy preserving data mining [1, 2].

## 2 Security Issues in e-Governance Systems

It is the key requirement of an e-Governance system with an interoperable secure infrastructure to meet the current and future needs. Such sub-system must act in coordination with its horizontally similar other sub-systems providing and getting information from varying sources [2].

Success of such a concept lies entirely in the hands of the citizens, their trust and confidence. Although, prone to security thefts, such a system can greatly introduce a revolutionary change in the public and private sector organizations.

The basic security requirements include [1]:
- Client authentication of the message and content verification.
- Sender and receiver authentication.
- No information, messages and data leaks.
- 100% availability and reliability.
- Confidentiality of the messages and overall system working.

The solutions to these any many more security issues can be to adopt existing and contemporary technology that ensures safety throughout the communication process. This can be done in combination with LDAP, PKI, PKI Smart Cards, SSO (Single Sign On), Web Single Sign On, Shibboleth etc… [1] The combination of these methods can help design efficient and secure e-governance architecture.

## 3 Privacy Issues in e-Governance Systems

Privacy of the information in a web-based e-governance system is equally important issue that cannot be compromised as the information of citizens on web is a valuable resource which needs its privacy and confidentiality. This data may be for person identification details, his family details, or any other personal information which is to be kept private. Also, as Government to citizen interaction increases, there is a need to safeguard the Government activities and data from information thefts [2]. In today's era, when personal information has become an extremely valuable resource, it is necessary to enforce security and legal restrictions to such hacking activities and data thefts.

For example, with the available personal information on web, whole profile of a person can be created, and depending on the confidentiality of hacked data, it can re-

________________________
- *Dr. Durgesh Kumar Mishra is with the Department of Computer Engineering Acropolis Institute of Technology and Research, Indore, India.*
- *Samiksha Shukla is with the Department of Computer Engineering, Christ University, Bangalore.*

sult in any kind of loss, which can be financial loss, password theft, account information loss, address and tax details stolen and so on. If bank information is lost, it can result to online banking theft.

Jurisdiction and a separate legal Department which can enforce this code of conduct of e-Governance need to be setup, to keep an eye on such malicious activities [1]. A web security Board should be setup to check and control the online thefts and other malicious activities that cause great information loss and thefts. Till now, in India we haven't yet been able to implement such strict web regulatory mechanism, but it has been successfully worked out in US [1].

As per the US privacy Legislation, there is a privacy certificate that is signed by citizens at the time they fill such privacy sensitive information in which they define the authentication and access levels which must be enforced on there data. Later, whenever this information is requested for web, first the certificate is checked for the validity of the access, and only if it is valid, it is sent to web, else access is denied. Similarly, there is other Government acts as well that define and control the privacy rights of the individuals [1].

## 4 Security Solutions to e-Governance

A single safety mechanism cannot suffice the security solution, but the complete communication must be must be kept safe using a combination of several security mechanisms. Some of the security methods are:

**4.1 Public Key Infrastructure:** It provides strong authe tication and secure communication to the entities involved. It is based on asymmetric keys and digital certificates to enable public key cryptography. It consists of a trusted third party called the certificate authority which binds public key to the entities involved. Certificate Authority maintains a table containing entries for each entity along with its public key and other entries. This infrastructure has various benefits like cost effective, interoperable, and consistent [1].

**4.2 Smart Cards/National Identity Cards (NIC):** It is hardware based cryptographic mechanism in which a card reader with the desired functionality and it implements the authentication and security mechanism. It too stores the digital certificates, private and public keys other entity related information. It performs the entire task with minimal human intervention. Additionally, these can also be used as electronic identification mechanism. Its benefits are convenient, strong authentication mechanism which is a one time investment [1].

Alternatively, NICs can also be employed with PKI to provide similar functionality.

**4.3 LDAP (Light Weight Directory Access Protocol):** It is an Internet Standard Protocol in which directories are organized according to X.500 data model. It can be used to issue, revoke (CRLs) and organize PKI certificates by means of directories [2].

**4.4 SSO (Single Sign On):** This mechanism allows user to sign in once and then make use of multiple resources with a single sign in [2].

**4.5 Web Single Sign on (SSO):** It allows surfing across or within organizational boundaries. By this, authorized decisions about a site for a particular entity pertaining to a particular entity can be made [2].

**4.6 Shibboleth:** It is a Security Assertion Mark up Language. Its key concepts include Federated Administration, Access Control Based on Attributes and Activity Management of Privacy. A collaboration of these above mentioned security architecture can give evolutionary Horizontal Infrastructure [2].

## 5 Privacy Solutions to e-Governance

Privacy is essentially an important right of each individual and must be protected by regulatory policies or legislations. It cannot be ensured without lawful acts and legal amendments. India, although has been a Democratic nation in which every Government activities have been fully transparent and well known, but now in this Internet Age, privacy of an individual is a major concern and has to be made through restricted access. Some of the privacy measures can be as under:

**5.1 Legal framework for Privacy Enforcement of Individual:** Prior this revolution, Government must frame out a strong set of principles and rules to work out the whole e-governance infrastructure. Some legal framework like US Policy Act of 1974 [2], the Policy Act of Senator, the Social Security Number Misuse Prevention Act, the Notification of Risk to Personal Data Act etc… must be prepared so that there are unambiguous and transparent rules for smooth running of the system across web. These rules can be similar to our fundamental rights and duties, which suggests that rights abide duties. Such acts must ensure that every citizen can insert, delete, modify, and view only his own record without any permission to access other records [2]. They must be able to determine what all records pertaining to them have been collected, allow purposeful access of record and so on. Also, the similar restrictions must be made for Government employees maintaining the record so that they may not purposely or mistakenly manipulate any record. Large fines and appropriate punishments for the hackers and malicious elements must also be defined so that such activities can be avoided or rather prohibited completely [2].

**5.2 Privacy Enhancing Technologies:** Privacy Rules enforcement is as important as its initial definition. Without its enforcement, everything shall be just a matter of trust of citizens which always very from person to person. Privacy Enhancing Policies are defined for this purpose to reinforce the policy rules from time to time. Several technologies exist that can address the privacy policies of the Government, Prominent among which are as under:

**5.3 Privacy Specification Language:** The e-Government Act of 2002 requires Federal Agencies to put in place the privacy protection of information collected electronically [2].

The World Web Consortium Standard, the Platform for Privacy Preferences or P3P, is a formal language for privacy communication to customers. It is a kind of specification that covers all the access and manipulation permissions and authorization levels. Each transaction on web, before being executed is checked for validity with the help of these specifications, and if these transactions are found valid they are executed or else aborted as invalid [2].

Another example of such a language which provides machine enforceable policies is IBM's Enterprise Privacy Authorization Language [EPAL]. It is an XML based Privacy Specification Language [2].

5.4    **Privacy during Data Mining:** In order to avoid false profiles, hacking and information thefts during the information retrievals, it is required. Such methods may include data modification, randomization, encryption, selective transformation, perturbation etc… [2].

Another solution to all e-Governance problems is Secure Multiparty Computation which shall be discussed in the next section.

5.5    **Privacy Preserving Databases:** Several advanced databases provide all the privacy related constraints to be specified on the data. Such mechanisms may include strong authentication, single sign on, LBAC policies, encryption schemes for databases, virtual private databases, role based authentication, Hippocratic databases etc… These databases allow mining process to take place only if they don't impose any privacy and security threats, and thus, ensures safe working [2].

**5.6 Transactional Privacy:** One way of achieving transactional privacy is by encryption. The transactional data is supplied in encrypted form and once the transaction is complete, it is again encrypted. So, even if the hacker hacks the data, he gets nothing but only garbage string. Alternatively, a direct encrypted data connection can also be established in which encryption is managed automatically [2].

Besides this, a special program can also be used that runs in background to safeguard all the ongoing transactions. An Administrator can also be employed which take cares of all the authorization checks and transactional validity [1].

5.7    **Statistical Data Protection:** Selective disclosure of statistical data is another safe approach in which statistical estimates are disclosed publicly without revealing the information or identity of an individual. The various ways in which such statistical Disclosure Control [SDC] can be implemented are query restriction, data perturbation, output perturbation. Similar to SDC are some other database protection techniques like tabular data protection, dynamic databases, microdata protection, data anonymization and anonymized data analysis, use of privacy brokers for proper privacy tracking and many more [2].

# 6  Secure Multiparty Computation [SMC] as a Comprehensive Solution to Privacy and Security Issues of e-Governance

The importance of data in transactional and computational environment cannot be overlooked as its loss can cause great chaos and problems [7]. When thinking of e-Governance infrastructure in a web based environment, we have to take into consideration some potential threats in that environment and their solutions or avoidance mechanisms [9]. Internet is a vast distributed architecture. Therefore, transactions and computations need to be designed with this architecture in mind. Secure Multiparty Computation provides a safe and efficient distributed computing environment in which no safety requirements of individuals are compromised [10]. Instead, it provides such an environment in which the privacy and confidentiality of the system as well as of the system is promised in polynomial time complexity and no extra overheads. At present, it is one of the unbeatable solutions to security.

The aim of a secure multiparty computation task is for the participating parties to securely compute some function of their distributed and private inputs [10]. In SMC, several parties or workstation who wish to compute some results or carry out certain transaction, send their inputs to some trusted third party [TTP] which then computes the result and then announces the results publicly [10]. In this scenario, each party learns nothing more than their own inputs and the final results. Also, the main requirements of a transaction, i.e. Correctness, independence of inputs/outputs, fairness and privacy are guaranteed [8].

# 7  Multi-agent Secure Multiparty Computation using Arithmetic Cryptography

Agent may be defined as a TTP that does the computation for the distributed parties. A multi-agent scenario basically consists of multiple agents that carry out computation, both in ideal and adversial manner [5]. For each task, arbitrary numbers of agents are employed, and the computations are made. A result is said to be correct if at least n/3 agents return the same result [5]. Moreover, arithmetic cryptography helps to secure data/ information leaks as the whole data on the channel and at the agents are full proof encrypted and thus, cannot be interpreted [12]. Even if hacked, it is of no use. Any of the protocol can be used by these agents for the computations as long as they yield the final results correct. The Architecture of Multi-agent SMC is given in fig.1.

In this architecture, Decision makers are the trustworthy components that check the agent's availability and trust and accordingly allocate tasks. The data channels are encrypted data channels in which data travels in encrypted manner. Parties fragment the data and then send it along the encrypted data channels to the decision makers for whatever task like, some mining task, statistical analysis, mathematical computation or else. These fragments are sent to the decision makers which provide some interme-diate conclusions according to the available data and requested operation. These intermediate conclusions are then forwarded to the agents that make the final concluding remarks and the result from the majority of agents is taken to be correct.

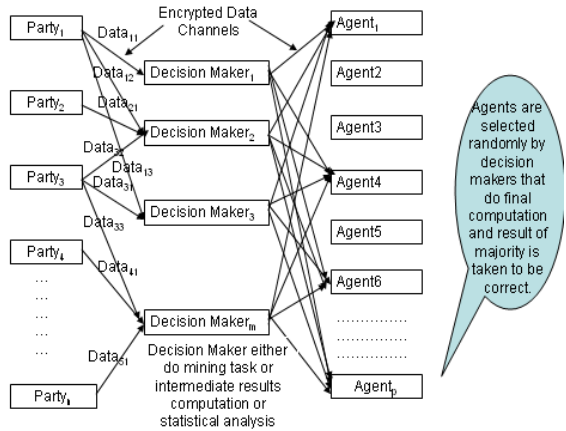

Fig. 1 Architecture of -agent Secure Multiparty Computation

As the data from parties is encrypted, the chances of being hacked are very less and moreover, decision makers contain the agent's past transactional performance and it selects the new agent as per its performance. The correctness of result comes from the strength of the cryptographic algorithms used. Special arithmetic encryption algorithms are designed for the purpose which provides reliable encryption. Agents are also not allowed to decrypt the data, but they apply the computations on same encrypted data. Thus the threat of agents being unreliable is also completely removed. Moreover, if some agent becomes malicious and does false computation or gives wrong results, it can be caught immediately, as we are using multiple agents, so the results from their majority will still be correct. Thus, the above mentioned architecture confirms all the requirements of secure e-governance and provides all the provisions under one architecture, which, as we have seen above are found very distributed through other mechanisms. We have to use their combinations to achieve desired features, whereas, here we get all of it under one unified architecture.

## 8 Features of the Multi-agent Secure Multi-party Computation

The protocol as we discussed above meets the safety requirements of our e-Governance System through its various features which are as under:

**8.1** **Security**: The security is guaranteed by its three level architecture. Data is sent over encrypted and does not move collectively; instead it is fragmented so that no one can get the complete data. Also, the neither the decision makers nor the agents untrustworthiness can affect the security of the system as they have just mere sequence of encrypted bits which gives no sense and is useless.

**8.2** **Privacy**: The decision makers helps in maintaining privacy as the data packets cannot be identified as to which party they belong. They provides an abstraction in between the agents and parties, which otherwise would have been linked. The agents selected for computation are also not revealed by the decision makers. If in worst case, if they even disclose the agent, no one can affect the data as it if encrypted and also agents don't know its decryption mechanism.

**8.3** **Correctness**: The correctness comes from the fact correctness of the computation procedure that has been used which is completely robust. It always guarantees correct results. Also, as the result is taken by the honest majority and not by a single agent, there is no possibility for incorrect results.

## 9 Performance Analysis of the Multi-agent Secure Multiparty Computation

The data is forwarded to decision makers in encrypted form. Therefore, additional encryption mechanism needs to be implemented. Let the total number of decision makers be $D_n$. Since, the data is first fragmented and they forwarded, hence, the probability of hacking the data of $n^{th}$ party with r data fragments is,

$$\Pr(n)_{data} = 1/r \qquad (1)$$

This probability of data hacking increases as the number of packets r increases as shown in fig. 2.
Also, since there are m decision makers, then the probability of decision maker to be corrupt is,

$$P_m(n)_{decision\ maker} = 1/m \qquad (2)$$

Therefore, the probability of the decision maker to become bias and make wrong decisions can be observed as shown in fig. 3.

Also, the numbers of agents are p; hence the probability of selecting a wrong agent by a biased decision maker is given by,
$$P(n)_{wrong\ agent} = (1/m) + (1/p) - (1/(m \times p))$$

Although, the number of agents and decision makers may vary, but the decision makers are always less than the agents. Therefore, the probability of this situation also decreases as the number of agents increases, and it even becomes better if we increase the number of decision makers also with increase in number of agents.

## 10 Conclusion

Thus, from the above analysis, we can observe that secure multiparty computation can be an effective solution to e-Governance issues and helps to solve is major issues without much complexity.

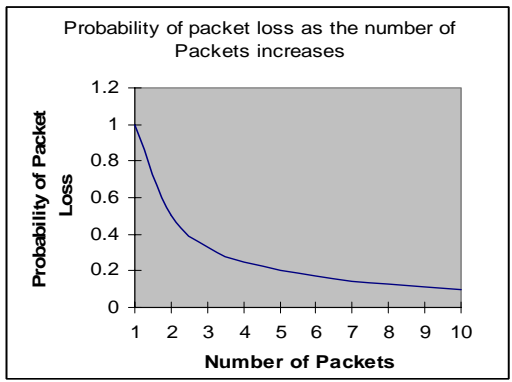

Fig. 2. Probability of packet loss with increase in number of packets.

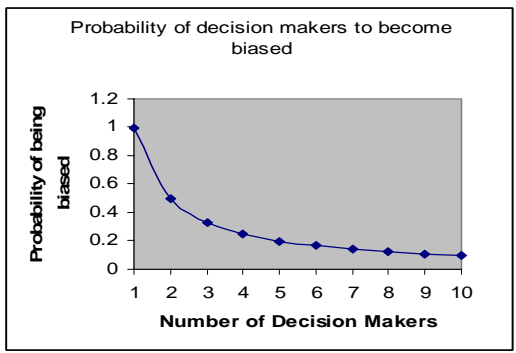

Fig. 3 Probability of the decision makers to become biased with increase in number of Decision Makers

## REFERENCES

1. Claudio Biancalana and Francesco Saverio Profiti "Security and Privacy Preserving Data in e-Government Integration", http://www.esiig2.it , Department of Computer Science and Automation, Roma Tre University, Rome, Italy.
2. Jaijit Bhattacharya," Privacy Technology for E-Governance", Department of Management Studiesn Indian Institute of Technology, Delhi, India, pp: 114-124.
3. Tu Bao Hu,"Privacy Preserving Data Mining and E-Commerce and E-Governance", School of Knowledge Science, Japan Advanced Institute of Science and Technology and IOIT, Vietnamese Academy of Science and Technology.
4. Arry Brandt, Lois Delcambrie, Sharon Dawes, Howard Bradsher-Fredrick "Being Successful in Digital Government Project", Birds of a Feather Session, University of Wisconsin-Madison, 2004.
5. Jacques Calmet, Regine Endsuleit, Pierre Maret "A Multi Agent Model for Secure and Scalable E-Business Transactions", http://www.avalon.ira.uka.de.
6. Jacques Calmet, Regine Endsuleit, "An Agent Framework for Legal Validations of E-Transactions", University of Karlsruhe, Germany.
7. Rachet GreenStadt, "Privatizing Constraint Optimization", Harvard University.
8. Durgesh Kumar Mishra and Manohar Chandwani, "A Zero Hacking Protocol for Secure Multiparty Computation using Multiple TTP", in the proceeding of Tencon'08, 19-21 Nov. 2008, pp:1-6.
9. Y.Lindell and B. Pinkas, "Privacy Preserving Data Mining". In advances in Cryptography-CRYPTO-2000, pp 36-54, Springer-Verlag, August 24 2000.
10. O. Goldreich, "Secure Multiparty Computation", September 1998 [Working draft] Online available on: http://www.wisdom.weizmann.ac.il/~oded/pp.html.
11. Vassilios S. Verykios, Elisa Bertino, Igor Nai Fovino, Loredana Parsiliti Provenza, Yucel Saygin, Yannis Theodoridis, "State-of-The-Art in Privacy Preserving Data Mining", SIGMOD Record, Vol. 33, No. 1, March 2004.
12. Durgesh Kumar Mishra; Manohar Chandwani,"Arithmetic cryptography protocol for secure multi-party computation", in the proceeding of SECON'07, 22-25 March 2007 pp:22 – 22.

*Authors Profile*

*Dr. Durgesh Kumar Mishra*
Professor (CSE) and Dean (R&D), Secretary IEEE MP-Subsection, Acropolis Institute of Technology and Research, Indore, MP, India, Ph - +91 9826047547, +91-731-4730038

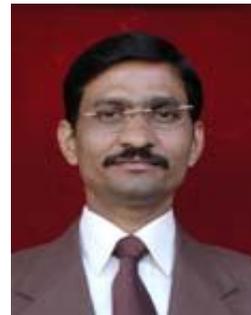

**Biography**: Dr. Durgesh Kumar Mishra has received M.Tech. degree in Computer Science from DAVV, Indore in 1994 and PhD degree in Computer Engineering in 2008. Presently he is working as Professor (CSE) and Dean (R&D) in Acropolis Institute of Technology and Research, Indore, MP, India. He is having around 20 Yrs of teaching experience and more than 5 Yrs of research experience. He has completed his research work with Dr. M. Chandwani, Director, IET-DAVV Indore, MP, India in Secure Multi- Party Computation. He has published more than 60 papers in refereed International/National Journal and Conference including IEEE, ACM etc. He is a senior member of IEEE and Secretary of IEEE MP-Subsection under the Bombay Section, India. Dr. Mishra has delivered his tutorials in IEEE International conferences in India as well as other countries also. He is also the programme committee member of several International conferences. He visited and delivered his invited talk in Taiwan, Bangladesh, USA, UK etc in Secure Multi-Party Computation of Information Security. He is an author of one book also. He is also the reviewer of tree International Journal of Information Security. He is a Chief Editor of Journal of Technology and Engineering Sciences. He has been a consultant to industries and Government organization like Sale tax and Labor Department of Government of Madhya Pradesh, India.

*Samiksha Shukla*
Asst. Professor (CSE),
Christ University, Banglore, India.

**Biography**: Samiksha Shukla has received M.Tech. degree in Computer Science from DAVV, Indore in 2005. Presently she is working as Asst. Professor (CSE) Christ University, Banglore, India.. She is having around 04 Yrs of teaching experience. She is doing her research work with Dr. Durgesh Kumar Mishra.